\pdfoutput=1

\documentclass[final,3p,times,square]{elsarticle}

\usepackage{graphicx}%
\usepackage{multirow}%
\usepackage{amsmath,amssymb,amsfonts}%
\usepackage{amsthm}%
\usepackage{mathrsfs}%
\usepackage[title]{appendix}%
\usepackage{xcolor}%
\usepackage{textcomp}%
\usepackage{manyfoot}%
\usepackage{booktabs}%
\usepackage{algorithm}%
\usepackage{algorithmicx}%
\usepackage{algpseudocode}%
\usepackage{listings}%
\usepackage{amssymb}
\usepackage{amsmath}
\usepackage{graphicx}
\usepackage{dcolumn}
\usepackage{bm}
\usepackage[utf8]{inputenc}
\usepackage[T1]{fontenc}
\usepackage{etoolbox}
\usepackage{booktabs} 
\usepackage{array} 
\usepackage{multirow} 
\usepackage{makecell} 
\usepackage{threeparttable}
\usepackage{tabularx}
\usepackage{float} 
\usepackage{url}
\usepackage{geometry}
\usepackage[markup=underlined]{changes} 
\usepackage{lineno}

\journal{Nonlinear Dynamics}

\begin{document}

\begin{frontmatter}


\title{
A Data-Driven Framework for Discovering Fractional Differential Equations in Complex Systems
}


\author[1,2]{Xiangnan Yu}
\author[1]{Hao Xu}
\author[3]{Zhiping Mao}
\author[4]{Hongguang Sun}
\author[5]{Yong Zhang}
\author[2]{Zhibo Chen}
\author[1]{Dongxiao Zhang}
\author[1,6]{Yuntian Chen\corref{cor1}}

\address[1]{Ningbo Key Laboratory of Advanced Manufacturing Simulation, Eastern Institute of Technology, Ningbo, Zhejiang 315200, China}
\address[2]{Department of Electronic Engineer and Information Science, University of Science and Technology of China, Hefei, Anhui, 230026, China}
\address[3]{School of Mathematical Sciences, Eastern Institute of Technology-Ningbo, China}
\address[4]{Institute of Hydraulics and Fluid Mechanics, Hohai University, Nanjing 211100, China.}
\address[5]{Department of Geological Sciences, University of Alabama, Tuscaloosa, AL, USA.}
\address[6]{Zhejiang Key Laboratory of Industrial Intelligence and Digital Twin, Eastern Institute of Technology, Ningbo, Zhejiang 315200, China}
\cortext[cor1]{Author for correspondence. Email-address: ychen@eitech.edu.cn (Yuntian Chen)}

\begin{abstract}


In complex physical systems, conventional differential equations fall short in capturing non-local and memory effects. Fractional differential equations (FDEs) effectively model long-range interactions with fewer parameters. However, deriving FDEs from physical principles remains a significant challenge.  This study introduces a stepwise data-driven framework to discover explicit expressions of FDEs directly from data. The proposed framework combines deep neural networks for data reconstruction and automatic differentiation with Gauss-Jacobi quadrature for fractional derivative approximation, effectively handling singularities while achieving fast, high-precision computations across large temporal/spatial scales. To optimize both linear coefficients and the nonlinear fractional orders, we employ an alternating optimization approach that combines sparse regression with global optimization techniques. We validate the framework on various datasets, including synthetic anomalous diffusion data, experimental data on the creep behavior of frozen soils, and single-particle trajectories modeled by L\'{e}vy motion. Results demonstrate the framework's robustness in identifying FDE structures  across diverse noise levels and its ability to capture integer-order dynamics, offering a flexible approach for modeling memory effects in complex systems.







\end{abstract}

\begin{keyword}
Fractional differential equations; Knowledge discovery; Sparse regression; Gauss-Jacobi quadrature; Machine learning.
\end{keyword}
\end{frontmatter}

\section{Introduction}

Differential equations are fundamental tools for modeling a wide range of physical phenomena, from solid mechanics to fluid dynamics. These equations encode physical laws, such as Newton's laws of motion and conservation laws, into mathematical formulation. Nevertheless, conventional model development through first-principles-based approaches faces challenges, particularly in complex systems characterized by the interplay of multiscale phenomena. The model construction for such systems proves challenging due to uncertainties in system characterization, often resulting in oversimplified representations that inadequately capture memory effects and nonlinear dynamics.
Benefiting from growing data acquisition and storage capabilities has significantly improved the utilization of natural system data, enabling data-driven models to predict complex phenomena \citep{reichstein2019deep}. However, most machine learning models act as "black boxes" with limited physical interpretability \citep{yang2024data}. Physics-informed neural networks (PINNs) \cite{raissi2019physics} integrate known physical laws directly into the lost function of neural networks, providing clear physical interpretation. However, the prerequisite physical knowledge (i.e., governing equations) required by PINNs remains unresolved.

 A new paradigm in equation construction through data-driven methodologies \citep{brunton2024promising,xu2024uncovering} overcomes these limitations. This strategy circumvents traditional derivation challenges by directly extracting governing equations from observations, bypassing many limitations inherent in conventional first-principles derivations that rely on explicit physical law formulations and mathematical simplifications. Meanwhile, the outcome differential equation enables the extrapolated prediction. 
 A pioneering method in this field is symbolic regression \citep{koza1994genetic}, which uses an evolutionary algorithm to find optimal combinations of coefficients and candidate analytical equations, balancing model simplicity with accuracy. Based on symbolic regression, Bongard and Lipson \cite{bongard2007automated} introduced a technique to learn coupled nonlinear ordinary differential equations directly from time series, while Schmidt and Lipson \cite{schmidt2009distilling} expanded this to derive conservation laws in dynamic systems from motion-tracking data. Sparse regression is another important framework for identifying structural forms of differential equations, Brunton et al. \cite{brunton2016discovering}  proposed the SINDy framework, a sparse regression method for deriving the governing equation from noisy data, particularly, through dimensionality reduction for partial differential equation extraction. However, SINDy struggles with high-dimensional data, limiting its applicability, as many experiments collect observations in high-dimensional space (e.g., tracer concentration time-series at various locations in dispersive transport studies).
 Rudy et al. \cite{rudy2017data} extended SINDy to handle spatio-temporal synthetic data via STRidge, a sequential threshold ridge regression approach that constructs a candidate library of time and space derivatives and employs sparse regression with $l_0$-regularization. Chang and Zhang \cite{chang2019machine} developed a framework using the least absolute shrinkage and selection operator (LASSO) to learn subsurface flow equations, though both STRidge and LASSO rely heavily on accurate numerical derivative approximations, making them  sensitive to data sparsity and measurement noise. To address these challenges,  Messenger and Bortz \cite{messenger2021weak} extended SINDy to Weak-SINDy (WSINDy) by applying a convolutional weak formulation, successfully uncovering PDEs from noisy data. Fasel et al. \cite{fasel2022ensemble} further proposed an ensemble-SINDy (ESINDy) which applies statistical ensemble to enhance WSINDy's robustness with noisy datasets. Omar et al. \cite{omar2024robust} proposed a physics-informed data-driven approach for identifying the nonlinear dynamics from experimental data with measurement noise. Zhang et al. \cite{zhang2021discovering} discovered governing equation from data under white noise, by transforming the ambient noise into stochastic equation. While these approaches have improved robustness, they lack data interpolation capability, limiting their applicability to non-uniformly sampled data. This fundamental limitation has motivated the adoption of deep neural networks (DNNs), which serve as surrogates producing reconstructed data with reduced noise and enhanced (in arbitrary) resolution. Based on DNNs, Xu et al. \cite{xu2019dl} combines them with STRidge to discover conventional PDEs under noisy and sparse data, with the proposed method achieving better results than the original STRidge. Both et al. \cite{both2021deepmod} proposed a deep learning-based LASSO to enhance robustness in model discovery with noisy, sparse data. Raissi et al. \cite{raissi2019physics} utilized PINNs to study inverse problems of PDEs under sparse and noisy data conditions, though this method requires prior knowledge of the PDE structure. It is important to note that studies on discovering governing equations can be broadly categorized into two types: structure discovery and coefficient discovery \cite{chen2022integration}, PINNs belong to the latter, requiring substantial prior knowledge of PDE structure, while the other methods introduced above fall under the former.
Xu and Zhang \cite{xu2021robust} proposed the R-DLGA framework, which combines PINNs with genetic algorithms to enable the discovery of PDEs without prior knowledge of their PDE structure, achieving to resilience to noise levels up to 50\%.   
 
 However, the frameworks mentioned above are limited to recovering basic governing equations in simple system. In contrast, natural systems are typically complex, characterized by multiscale physics and anomalous phenomena. For example, dispersive transport in heterogeneous porous media exhibits anomalous dispersion, characterized by non-Gaussian spatial concentration distribution or nonlinear mean squared displacement growth \citep{metzler2000random, comolli2019mechanisms}. To capture such anomalous phenomena, an alternative method is direct modeling through tradition models with time- and space-variable parameters \cite{rudy2019data}, such as using the advection-dispersion equation with fine-resolution hydraulic conductivity fields to model solute transport in heterogeneous media \citep{wheatcraft1988explanation, salamon2006review}. However, this approach requires detailed parameterization of aquifer heterogeneity, which poses significant challenges for the aforementioned data-driven discovery methods. To derive parametric PDEs without prior knowledge, more advanced model discovery algorithms are needed. Rudy et al. \cite{rudy2019data} improved STRidge to Group STRidge, showing that it outperforms other frameworks for discovering parametric PDEs. Xu et al. \cite{xu2021deep} used integral forms of PDEs to mitigate noise, enabling the identification of PDEs with heterogeneous parameters. Chen et al. \cite{chen2022symbolic} proposed SGA-PDE, which combines symbolic regression with genetic algorithms to achieve the flexible representation of some parametric PDEs. Additionally, SGA-PDE is an open-form equation discovery framework, meaning it does not rely on the candidate library. Sun et al. \cite{sun2024data} employed SGA to discover KdV equation and nonlinear Schrödinger equations by multi-soliton solutions. Du et al. \cite{du2024discover} integrated reinforcement learning into the SGA-PDE framework to improve the accuracy and efficiency of discovering PDEs with fractional structure (i.e., variable fraction parameters) and higher-order derivatives. Furthermore, they introduced a robust version of the framework to handle highly noisy data \citep{du2023physics} and proposed a Large Language Model (LLM)-guided equation discovery framework to break through barriers to interdisciplinary research \citep{du2024llm4ed}.

 Nevertheless, challenges remain: first, identifying heterogeneous parameters is time-intensive due to the large number of coefficients \citep{tartakovsky2020physics}. Second, issues such as mathematical ill-posedness and parameter equifinality, where different parameter sets yield similar transport behaviors \citep{beven2001equifinality}, complicate inverse modeling and can lead to incorrect models. Current approaches are limited to discovering PDEs with weakly varying parameters \citep{chen2022integration}. In natural systems, however, parameters often exhibit multiscale variability and may change significantly across both time and space, as seen in hydraulic conductivity fields in aquifers \citep{adams1992field}. Consequently, the applicability of current model discovery methods remains constrained in the context of highly complex systems.

 An intrinsic approach for reducing the difficulty of model discovery in highly complex systems is model upscaling, which reduces the degrees of freedom in the equations and the number of parameters using techniques such as mathematical homogenization or stochastic methods \citep{cushman2002primer}. A representative application of upscaling tools is dispersive transport in aquifers. Various parsimonious (upscaled) models have been developed to capture macroscopic transport in aquifers \citep{neuman1990universal, cushman2002primer}, including the stochastic model \citep{dagan1987theory}, continuous-time random walks \citep{berkowitz2006modeling}, fractional advection-diffusion equations \citep{benson2000application}, and the multi-rate mass transfer model \citep{haggerty1995multiple}. These models effectively capture dispersive transport in highly heterogeneous media with only a few additional parameters, significantly reducing the degrees of freedom and parameters of equations. 
Among the upscaled models mentioned above, fractional advection-diffusion equations are especially promising due to the well-studied mathematics of fractional calculus \citep{kilbas2006theory, podlubny1999fractional, meerschaert2019stochastic} and various numerical algorithms \citep{guo2015fractional}. Beyond geology, fractional differential equations (FDEs) have been widely applied to model non-local dynamics in complex systems over the past several decades \citep{sun2018new}, including anomalous diffusion \citep{metzler2000random}, non-Newtonian fluids \citep{sun2018space}, creep and relaxation \citep{schiessel1995generalized} and continuous finance \citep{scalas2000fractional}, among others. 
Promoting the development of models that incorporate fractional calculus remains a significant effort \citep{sun2018new}.  Data-driven discovery of FDEs is essential to accelerate this development, which motivate this study.  However, discovering FDEs presents computational challenges. First, the singular nature of the power-law kernel function affects the accuracy and stability of solutions \citep{yang2022using}. Second, the nonlocality of convolution, with interactions over distance or time, necessitates the computation of fractional derivatives using the values of functions at multiple nonlocal nodes. Third, current sparse regression methods can only linearly optimize the sparse coefficients of candidate derivative terms, limiting their ability to adjust fractional orders, which are nonlinear parameters. Recent advancements in discovering FDE coefficients have been extensively explored using methods like  fPINNs \cite{pang2019fpinns} and Gaussian processes \cite{gulian2019machine}. However, these approaches are heavily dependent on prior knowledge of the equation structures. Neural fractional differential equations (Neural FDEs) \cite{coelho2024tracing} can model memory-dependent dynamics without requiring prior knowledge, and demonstrate strong extrapolation capabilities \cite{coelho2025neural}. However, their non-interpretable nature prevents the determination of explicit FDE forms. Since deriving explicit FDE expressions is crucial for understanding both the physical mechanisms and mathematical properties of these systems, developing white-box algorithms capable of extracting symbolic representations becomes fundamentally important. Recent works by \cite{singh2022data,vats2024new} has developed a white-box frameworks for extracting explicit expression of FDEs by addressing key challenges related to the optimization of fractional orders and preventing reliance on prior knowledge. However, their mesh-based derivative approximation approach suffers from two  limitations: (i) low computational efficiency at large scales due to dense spatiotemporal discretization requirements, (ii) insufficient accuracy due to unresolved singularity issues. Additionally, a broader challenge persists across data-driven PDEs discovery paradigms: insufficient robustness against measurement noise, which remains unaddressed in current methodologies.

The primary objective of this study is to develop a robust, efficient, and accurate framework. We propose a stepwise framework for discovering FDEs that effectively addresses these challenges. 
This paper makes several key contributions:
\begin{itemize}
    \item {Robust derivation of explicit FDE expressions from sparse, noisy data through DNN data reconstruction;}
    \item {Implementation of G-J quadrature  that simultaneously addresses singularity issues in fractional derivatives \citep{yang2022using} while outperforming mesh-dependent discretization methods in computational efficiency and accuracy \citep{pang2013gauss};}
    \item {Proposing an alternating optimization framework for systems with mixed linear and nonlinear coefficients (e.g., sparse coefficients and fractional orders, respectively).} 
    \end{itemize}

The remainder of this work is organized as follows: Section 2 presents the FDEs discovery framework, including definitions of fractional calculus and the stepwise method for discovering FDEs. Section 3 demonstrates the effectiveness of the proposed algorithm through various FDE discoveries. Section 4 discusses the characteristics of our method. Finally, Section 5 summarizes the conclusions and directions for future work.

\section{FDEs discovery framework}

\subsection{Fractional differential equation}\label{sec:frac}
We consider a general form of a partial differential equation extended by incorporating fractional derivatives as follows:

\begin{equation} \label{eq:fde}
    u_t^{(\alpha)}  = {\cal F}\left( {u, u_x, u_x^{(\beta)}, u_{xx} ,...} \right)\boldsymbol \xi,
\end{equation}
where the subscripts $t$ and $x$ denote the derivative of the function $u$ with respect to time and space, respectively. The superscripts "$(\alpha)$" and "$(\beta)$" represent the fractional order of differentiation in time and space, $\mathcal{F}(\cdot)$ is a linear operator to be determined, formed by the linear combination of the coefficient vector $\boldsymbol \xi$ and the various derivatives of $u$.


In engineering fields, fractional calculus is commonly defined in two forms: the Riemann-Liouville (R-L) definition and the Caputo definition. The R-L derivative is expressed as

\begin{equation}\label{eq:rl_def}
    u_x^{(\gamma)} = {}_{ - \infty }^{RL}D_x^\gamma u(x) = \frac{{{d ^n}}}{{d {x^n}}}{I^{n - \gamma }}u(x),
\end{equation}
and the Caputo derivative is given by
\begin{equation}\label{eq:cap_def}
    u_t^{(\gamma)} = {}_{ 0 }^CD_t^\gamma  u(t) = {I^{n - \gamma}}\frac{{{d ^n}u(t)}}{{d {t^n}}},
\end{equation}
where $I^{\gamma}$ denotes the $\gamma$-th (arbitrary real number) order of fractional integrals,

\begin{equation}\label{eq:frac_int}
    I_a^\gamma u(t) = \frac{1}{{\Gamma (\gamma )}}\int_a^t {{{(t - \tau )}^{\gamma  - 1}}u(\tau )}d\tau,
\end{equation}
and $n$ represents the integer obtained by rounding up the value of $\alpha$. For fractional derivatives with respect to time, the Caputo definition is commonly used because it provides reasonable initial conditions \citep{podlubny1999fractional}. Conversely, space fractional derivatives are often defined by the R-L definition, given the common assumption of an infinite computational domain. There is a relationship between the two definitions:

\begin{equation} \label{eq:rl_caputo}
{}_a^CD_x^\gamma u(x) = {}_a^{RL}D_x^\gamma  {u(x) - \sum\limits_{i = 0}^{n - 1} {\frac{{{{(x - a)}^{i - \gamma }}}}{{\Gamma (i - \gamma  + 1)}}{u^{(i)}}(a)} } 
\end{equation}
where $a$ denotes the initial node, and $i$ represents the integer order of the derivative. As shown in Equation~\eqref{eq:rl_caputo}, when the computational region is sufficiently large and boundary conditions are minimal, which is common in natural systems, the R-L derivative can be approximated by the Caputo derivative \citep{kilbas2006theory}.

\subsection{Sparse regression for linear coefficients}
Assuming that the function $\mathcal{F}$ (see Equation \eqref{eq:fde}) is composed of a linear multiplication of coefficients and candidate derivative terms, Equation \eqref{eq:fde} can be discretized as 

\begin{equation} \label{eq:sr}
    \textbf{u}_t^{(\alpha)}  = \Theta(\beta) \cdot \boldsymbol{\xi},
\end{equation}
where $\textbf{u}$ is the data, $\Theta$ denotes the candidate library, containing all possible (linear or nonlinear) terms including fractional order $\beta$ with respect to space, expressed as

\begin{equation}
    \Theta(\beta)  = \left( {\begin{array}{*{20}{c}}
        1&{u({x_1},{t_1})}& \cdots &{u_x^{(\beta)} ({x_1},{t_1})}& \cdots \\
        1&{u({x_2},{t_1})}& \cdots &{u_x^{(\beta)} ({x_2},{t_1})}& \cdots \\
         \vdots & \vdots & \ddots & \vdots & \ddots \\
        1&{u({x_M},{t_1})}& \cdots &{u_x^{(\beta)} ({x_M},{t_1})}& \cdots \\
         \vdots & \vdots & \ddots & \vdots & \ddots \\
        1&{u({x_M},{t_N})}& \cdots &{u_x^{(\beta)} ({x_M},{t_N})}& \cdots 
        \end{array}} \right),
        \label{eq:Theta}
\end{equation}
 $\Theta$ and $\textbf{u}^{(\alpha)}_t$ should be calculated prior to solving the learning problem, which involves evaluating the sparse coefficient vector $\boldsymbol \xi$. The primary task of this study is searching the optimal parameters including $\alpha$, $\beta$ and $\boldsymbol{\xi}$ for fitting the data. Generally, linear least-squares regression is used to determine the optimal coefficients $\boldsymbol{\xi}$ by minimizing the residual function:

 \begin{equation} \label{eq:min}
    \boldsymbol{\xi } = \mathop {\arg \min }\limits_{\boldsymbol{\hat \xi}} \left( \left\| {{\bf{u}}_{t}^{({\alpha })} - {\Theta(\beta) \cdot \boldsymbol{\hat \xi}}} \right\|_2^2  \right).
 \end{equation}

However, although least-squares algorithm yields the mathematically optimal coefficients, it often results in trivial, unreasonable values of $\xi$ (due likely to overfitting), reducing the model parsimony and interpretability. Sparse regression addresses this issue by eliminating negligible coefficients \cite{brunton2016discovering}. A key feature of sparse regression is regularization term, which is added to the residual function to penalize the number of non-zero coefficients
\begin{equation} \label{eq:min_reg}
    \boldsymbol{\xi } = \mathop {\arg \min }\limits_{\boldsymbol{\hat \xi}} \left( \left\| {{\bf{u}}_{t}^{({\alpha })} - {\Theta(\beta) \cdot \boldsymbol{\hat \xi}}} \right\|_2^2  + \lambda \|\boldsymbol{\hat \xi}\|_0 \right),
\end{equation}
where $\lambda$ denotes the regularization parameter, and  $||.||_0$ denotes $\ell^0$-norm, which counts the number of non-zero elements in a vector. However, $\ell^0$-norm regularization is non-convex and leads to a computational complex optimization problem. In contrast, $\ell^1$-norm regularization, referred to as the least absolute shrinkage and selection operator (LASSO), serves as a convex relaxation of the $\ell^0$-norm regularization and is defined as
\begin{equation} \label{eq_reg1}
    \boldsymbol{\xi } = \mathop {\arg \min }\limits_{\boldsymbol{\hat \xi}} \left( \left\| {{\bf{u}}_{t}^{({\alpha })} - {\Theta(\beta) \cdot \boldsymbol{\hat \xi}}} \right\|_2^2  + \lambda \|\boldsymbol{\hat \xi}\|_1 \right),
\end{equation}
where $\|\boldsymbol{\hat \xi}\|_1$ is the $\ell^1$-norm of $\boldsymbol{\hat \xi}$. LASSO has been successfully used to identify the linear coefficients of PDEs \cite{chang2019machine,both2021deepmod}. However, empirical evidence suggests that LASSO performs poorly in the presence of multi-collinear coefficients \cite{rudy2017data}. To address this limitation, sequential threshold ridge regression (STRidge) \cite{rudy2017data} has been proposed as an alternative approach. STRidge combines the $\ell^2$-norm  regularization, referred to as ridge regression, which is defined as
\begin{equation} \label{eq:ridge}
    \boldsymbol{\xi } = \mathop {\arg \min }\limits_{\boldsymbol{\hat \xi}} \left( \left\| {{\bf{u}}_{t}^{({\alpha })} - {\Theta(\beta) \cdot \boldsymbol{\hat \xi}}} \right\|_2^2 + \lambda \|\boldsymbol{\hat \xi}\|_2^2 \right),
\end{equation}
where$ \|\boldsymbol{\hat \xi}\|_2$ is the $\ell^2$-norm of the vector $\boldsymbol{\hat \xi}$, STRidge is realized by combining ridge regression with a dynamically adjusted threshold selecting non-zero coefficients (this process involves $\ell^0$-norm, for details see \cite{rudy2017data}). STRidge has been proven effective and are integrated into various advanced PDE discovery frameworks \citep{xu2019dl,chen2021physics}. To ensure the best performance of model discovery, we employ STRidge for the estimation of linear coefficient vector $\boldsymbol{\xi}$, while the optimization of nonlinear coefficients is discussed in Section \ref{sec:nonlinear}.

\subsection{Generating the library for fractional derivative} \label{sec:generating_library}
 
The elements in Equation~\eqref{eq:Theta} should be predefined, with integer-order derivatives conveniently computed using automatic differentiation. While generating fractional derivative terms is complex, calculating fractional-order derivatives requires function values from both local and nonlocal regions due to the inherent nonlocality in their definitions: as demonstrated in Equations \eqref{eq:rl_def} and \eqref{eq:cap_def}, where the convolution integrals extend from the origin to distant regions.
Moreover, data sparsity and noise complicate the calculation of derivative terms. To address these challenges, we employ deep neural networks (DNN) to simultaneously denoise and perform super-resolution reconstruction of sparse data fields. DNN serves as a surrogate model, with their capabilities in data interpolation and noise smoothing, making them well-suited for handling sparse and noisy data. Moreover, the automatic differentiation embedded in DNN can calculate the integer-order derivative component in fractional derivatives with high precision and robustness. The dataset is processed with a DNN as follows:

\begin{equation}
    u(\textbf{z}) \approx \hat u(\textbf{z}) = {y_n}({y_{n - 1}}(...({y_2({y_1}}(\textbf{z}))))),
\end{equation}
where $u(\textbf{z})$ and $\hat{u}(\textbf{z})$ denote the original dataset and reconstructed data, respectively, $\textbf{z}=(x,y,z,t)$ is the coordinate vector in space and time, and
\begin{equation}
    {y_i}(\textbf{z}) = \sigma ({\boldsymbol{\theta} _i}\textbf{z} + {\textbf{b}_i}), i = 1,2,...n,
\end{equation}
where $n$ denotes the number of hidden layers, $\sigma$ is the chosen activation function, and $\boldsymbol{\theta}_i$ and $\textbf{b}_i$ represent the weights and bias in the $i$-th layer, respectively.  The cost function for the DNN is defined as
\begin{equation}
    \mathcal{L}\left( {z,\omega} \right) = \frac{1}{N}\sum\limits_{i = 1}^{{N}} {{(\hat {u}(z_i ) - {u}(z_i))^2}} 
    \label{eq:dn_cost}
\end{equation}
where $N$ denotes the number of measurements. The loss function $\mathcal{L}$ is used to train the DNN.
Once a high-quality reconstructed data field is created, following the idea introduced by \cite{xu2019dl}, the integer-derivative terms can be analytically estimated within the DNN by applying auto-differentiation.
However, because fractional calculus invalidates the standard chain rule, fractional derivative terms cannot be generated solely through automatic differentiation of the DNN solution $\hat{u}(\textbf{z})$. Mesh-based methods, such as the finite difference method and the Gr\"{u}nwald-Letnikov difference scheme, have been used to obtain fractional derivatives within the deep learning framework, as in previous work \citep{pang2019fpinns}. However, these mesh-based methods may suffer from singularity issues that affect accuracy and stability when approximating fractional derivatives, and their global nature requires numerous equidistant auxiliary points from origin, leading to high computational cost at large scales.
 As an alternative, we employ the Gauss-Jacobi (G-J) quadrature approach \citep{pang2013gauss}, which utilizes orthogonal polynomials  to efficiently and accurately approximate fractional integrals via coordinate transformation. This approach overcomes the above limitations by:
 (1) overcoming singularity-induced accuracy problems via localized spectral approximation, and
 (2) achieving high-precision results at large scales with a few number of fixed computational nodes (as detailed in Appendix \ref{sec:gj}).
According to Equations \eqref{eq:rl_def} and \eqref{eq:cap_def}, fractional derivatives combine integer-order derivatives and fractional integrals. Thus, by integrating auto-differentiations in deep learning framework such as PyTorch into the G-J quadrature, fractional derivative terms can be efficiently computed in a fixed number of auxiliary nodes (generated by DNN). The procedure for generating fractional derivatives (e.g., in time at $t=t_i$) is summarized as follows: first, initialize the input and calculate the approximate integer derivative component using automatic differentiation. Second, rewrite the formulation of the fractional derivative 
\begin{equation}
    {\left. {\hat u_t^{(\alpha )}(x,t)} \right|_{t = {t_i}}} = {\left. {\frac{{{\partial ^\alpha }\hat u(x,t)}}{{\partial {t^\alpha }}}} \right|_{t = {t_i}}} = \frac{1}{{\Gamma (n - \alpha )}}\int_0^{{t_i}} {{{\left. {{\tau ^{n - 1 - \alpha }}\frac{{{\partial ^n}\hat u(x,t)}}{{\partial {t^n}}}} \right|}_{t = {t_i} - \tau }}d\tau } 
    \label{eq:gj_tfrac}
\end{equation}
to match the G-J quadrature via the variable transformation $\tau = t_i(1+\zeta)/2$, we obtain
\begin{equation}
    {\left. {\hat u_t^{(\alpha )}(x,t)} \right|_{t = {t_i}}} = \frac{{{{({t_i}/2)}^{n - \alpha }}}}{{\Gamma (n - \alpha )}}{\int_{ - 1}^1 {\left. {{{(1 + \zeta )}^{n - 1 - \alpha }}\frac{{{\partial ^n}\hat u(x,t)}}{{\partial {t^n}}}} \right|} _{t = \frac{{{t_i}}}{2}(1 - \zeta )}}d\zeta  . \label{eq:gj_dtfrac}
\end{equation}

\begin{figure}[htb] 
    \centering
    \includegraphics[scale=0.55]{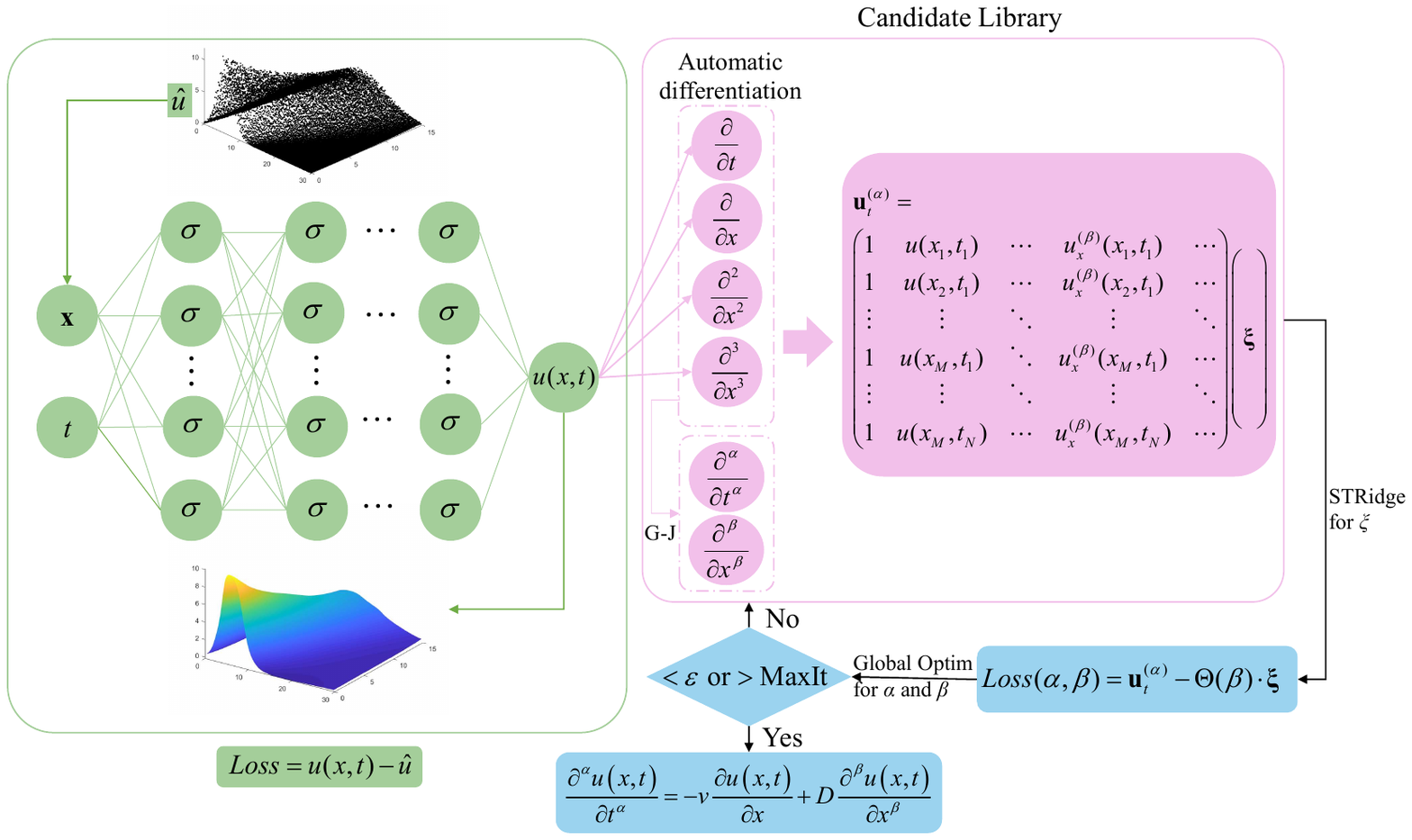}
    \caption{Workflow for data processing and generating the library of derivative terms. MaxIt in this figure represents the predefined maximum iteration number for optimization.}
    \label{fig:flow_lib}
\end{figure}

The calculation of derivative terms can be treated as an algebraic equation with a fixed number of terms, as shown in Equations \eqref{eq:gj_rule} and \eqref{eq:gj_frac}. 
In this study, we select five quadrature nodes; while more nodes can improve accuracy, five nodes generally offer a good balance between computational efficiency and accuracy. Fractional derivatives in terms of space follow the same process. The G-J quadrature outperforms mesh methods (e.g., finite difference method) for generating the library of derivative terms, since it requires only a few auxiliary nodes. Numerical examples are provided in \ref{sec:gj}. The workflow for generating the library of derivative terms is illustrated below (see Figure~\ref{fig:flow_lib}). 


\subsection{Stepwise optimization approach for nonlinear coefficients}\label{sec:nonlinear}

According to Equation~\eqref{eq:min}, the cost function comprises the mean squared error (MSE) related to the sparse vector $\boldsymbol \xi$, fractional orders $\alpha$ and $\beta$, and a regularization term, formulated as follows
  
 \begin{equation} \label{eq:reg}
    {\cal L(\alpha,\beta,\boldsymbol \xi)} = {\left\| {u_t^{(\alpha)}  - \Theta(\beta) \cdot \boldsymbol \xi } \right\|^2_2} + \lambda \left\| \boldsymbol \xi  \right\|^2_2.
 \end{equation}
 
 STRidge is employed to determine the linear coefficients $\xi$ of Equation~\eqref{eq:reg}. However, STRidge fall short in identifying FDEs due to the nonlocal nature of fractional derivatives, a process requiring simultaneous estimation of fractional orders (continuous nonlinear parameter \citep{xu2022discovery} embedded within the operator), alongside model structure discovery. Raissi \cite{raissi2018deep} proposed a framework using two DNNs to capture physical dynamics, effectively avoiding issues with nonlinear parameters. However, the resulting model is a black-box, lacking a closed form governing equation. The black-box nature hinders researchers from further studying the theoretical properties of governing equations.

\begin{figure}[htb]
    \centering
    \includegraphics[width=0.45\textwidth]{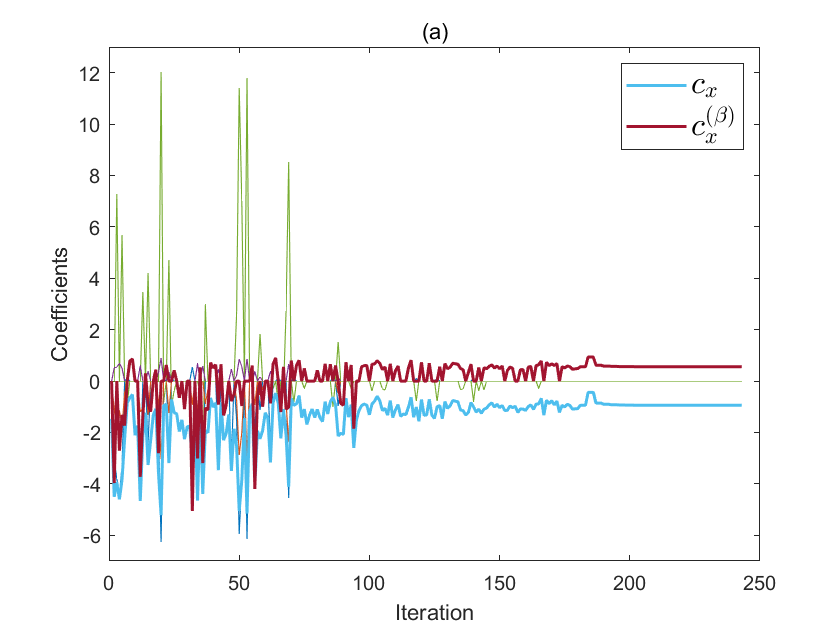}
    \includegraphics[width=0.45\textwidth]{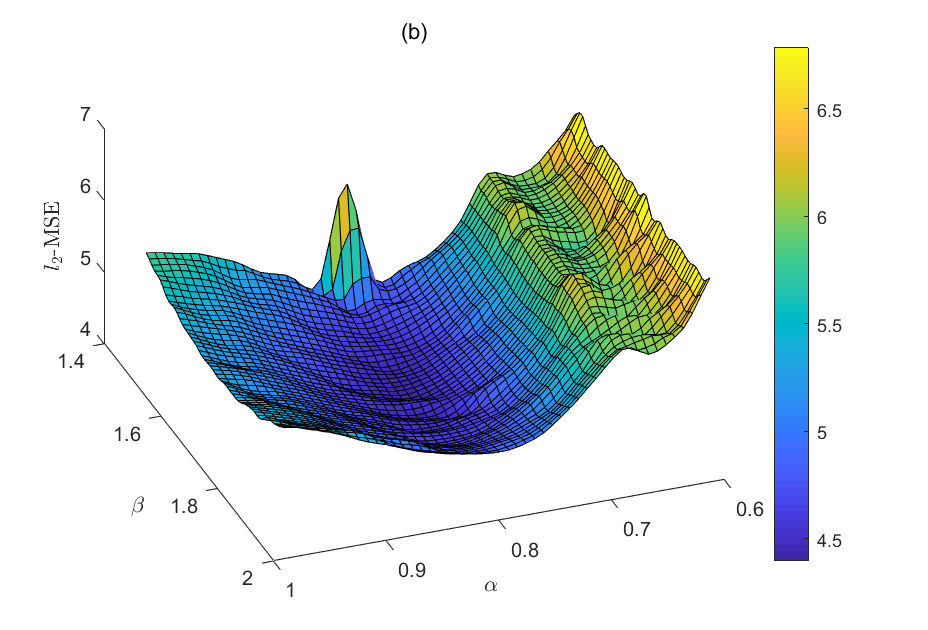}
    \caption{(a) Evolution of sparse coefficients estimated via our alternating optimization approach. The y-coordinate represents the best sparse coefficients $\xi$ estimated by STRidge under fixed fractional orders, the x-coordinate represents the iterations of the global optimization algorithm. (b) Evolution of the loss function~\eqref{eq:reg} with fractional orders. The specific example presented in this figure involves a fractional advection-diffusion equation that incorporates 5\% uniform noise, as discussed in Section~\ref{sec:fade_discovery}.}
    \label{fig:err}
\end{figure}

 To obtain both the optimal nonlinear fractional order and linear coefficients $\xi$, we design an alternating optimization approach that estimates $\xi$ and fractional orders sequentially (Figure~\ref{fig:err}a illustrates the alternating optimization procedure). Linear coefficients are determined by STRidge, while the fractional orders are estimated using a global optimization algorithm. Note that the loss function for optimizing fractional orders is non-convex and discontinuous due to the number of non-zero $\xi$ varies with fractional orders (see Figure~\ref{fig:err}b for details). In such cases, gradient-based optimization algorithms, such as the Adam algorithm and Newton's iterative methods, may encounter ill-posed problems at the discontinuous interface of the loss function. We evaluated several conventional global optimization algorithms, including Simulated Annealing (SA), Differential Evolution (DE), Powell Algorithm (PA), and Particle Swarm Optimization (PSO) for this task. Among these methods, DE and PA were respectively identified as the most effective approaches for multi-parameter and single-parameter optimization tasks. Therefore, DE is used in this study to estimate the fractional orders of fractional PDEs involving multiple fractional orders, while PA is used for fractional ODEs with a single fractional order. The optimization procedure consists of five main steps (listed in Algorithm~\ref{tab:pseudocode}). 
  Notably, Vats et al. \cite{vats2024new} also developed a sparse reconstruction-based approach for learning FDEs. The key difference between their work and ours lies in the methodology and scope of FDEs discovery from data. Vats et al. \cite{vats2024new} used mesh-based scheme to handle fractional derivatives, and used sparse reconstruction with LASSO and DE to identify fractional orders of time and space derivatives, applying this method to synthetic data and biological processes under conditions of low noise density. In contrast, our approach introduces a stepwise, data-driven framework that combines deep neural networks for data reconstruction and G-J quadrature to handle fractional derivatives. We optimize both sparse coefficients and fractional orders using a hybrid of sparse regression and global optimization algorithms. Utilizing the G-J quadrature's ability to compute fractional integrals with minimal quadrature nodes, our method demonstrates enhanced computational scalability. Due to the data reconstruction by DNNs, our framework offers superior robustness and flexibility in FDEs discovery, even under dense noisy conditions. Thus, while Vats et al. \cite{vats2024new} focused on sparse reconstruction for FDEs, our method integrates deep learning and advanced numerical techniques for more practical FDE identification in complex real-world systems.

 \begin{algorithm}
    \caption{Stepwise Approach for Discovery of FDEs}\label{tab:pseudocode}
    \begin{algorithmic}[1]
    \State \textbf{Step 1: Initialization}
    \State 1.1 Randomly initialize $\alpha = \alpha_0$ and $\beta = \beta_0$.
    \State 1.2 Generate the library using a deep-learning based method for $\alpha_0$ and $\beta_0$:
    \State \hspace{1.5em} $\mathbf{\Theta} \gets [\hat{u}_x^{(\beta_0)} = \sum_{i=1}^5 \omega_i \hat{u}_{xx}(\zeta_i,t),  
    \hat{u}_t^{(\alpha_0)} = \sum_{j=1}^5 \theta_j \hat{u}_t(x,\tau_j), \hat{u}_t, \hat{u}_x,...]$.
    \State
    \State \textbf{Step 2: Sparse regression to update linear coefficients}
    \State Estimate the optimal coefficient vector $\boldsymbol{\xi}$ using the STRidge algorithm:
    \State \hspace{1.5em} $\boldsymbol{\hat{\xi}} = \mathop{\arg \min}\limits_{\boldsymbol{\xi}} L(\alpha_0, \beta_0, \boldsymbol{\xi})$$+ \lambda \|\boldsymbol{\xi}\|_2^2$.
    \State
    \State \textbf{Step 3: Global optimization to update fractional orders}
    \State 3.1 Update $\alpha$ and $\beta$ using a global optimization algorithm:
    \State \hspace{1.5em} {$\{\hat \alpha, \hat \beta\} = \mathop{\arg \min}\limits_{\{\alpha, \beta\}} L(\alpha, \beta, \boldsymbol{\hat \xi})$}.
    \State 3.2 Rebuild the library based on the updated $\hat \alpha$ and $\hat \beta$: $\mathbf{\Theta} \gets [\hat{u}_x^{(\hat \beta)}\ |\ \hat{u}_t^{(\hat \alpha)}\ |\ \cdots]$.
    \State
    \State \textbf{Step 4: Alternating direction optimization iteration}
    \State  Repeat Steps 1 and 2 until the loss function $\cal L(\alpha,\beta,\boldsymbol \xi)$ converges to a minimum,
    \State \hspace{1.5em} or the maximum number of iterations is reached.
    \State
    \State \textbf{Output:} Return the optimized $\alpha$, $\beta$, and $\boldsymbol{\xi}$.
    \end{algorithmic}
    
    \end{algorithm}

\section{Results}
In this section, we validate the applicability of our method for identifying fractional differential equations directly from data, without requiring prior knowledge of the equation structure. We present a series of case studies, including synthetic data for the fractional advection-diffusion equation, experimental data on the creep process of frozen soils, and time-series data from single particle tracking that conforms to an $\alpha$-stable distribution. Initially, we consider a straightforward case study involving experimental data described by a fractional ordinary differential equation (ODE). The experimental data for this scenario is relatively straightforward to obtain. 

\subsection{Discovering the fractional Kelvin model from experimental data}

\begin{figure}[htb]
    \centering
    \includegraphics[width=0.45\textwidth]{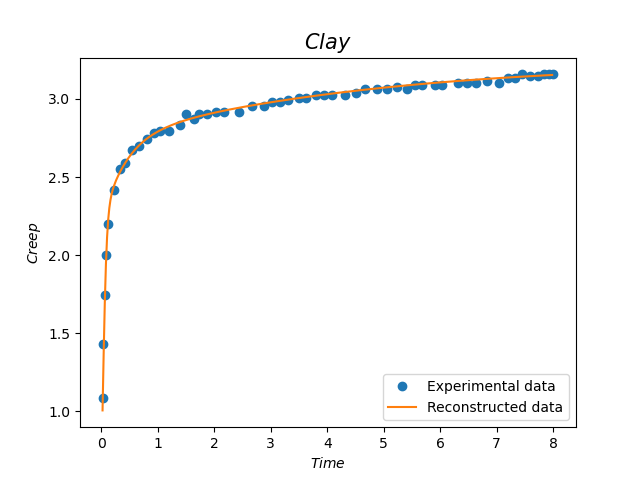}
    \includegraphics[width=0.45\textwidth]{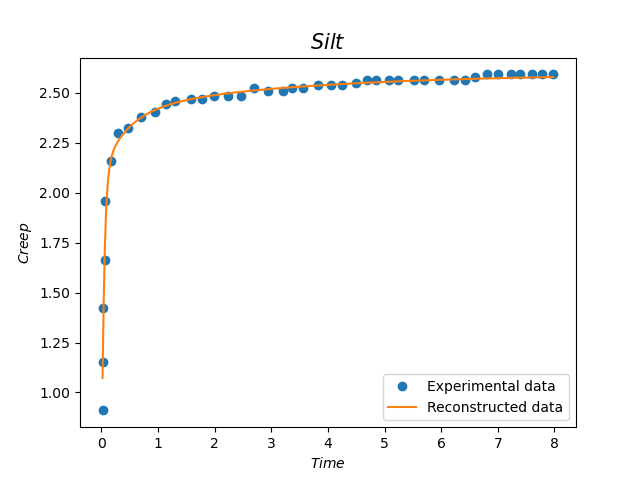}
    \caption{Reconstruction of experimental data for clay and silt.}
    \label{fig:creep}
\end{figure}

We examine the uniaxial compression creep of frozen soil as a case study. This creep process is complex, influenced by mechanical properties that include multiphase components and various environmental conditions. Traditionally, it has been modeled using ODEs to describe stress-strain relationships and incorporate memory effects.
Studies show that fractional constitutive models better capture the time-dependent behavior and memory effects in the creep process \citep{schiessel1995generalized}.
Here, we applied our method to derive the constitutive relationship (expressed by ODEs) between stress and strain in frozen soil, using data from a coal mine in Huainan, China, measured by \cite{chen2013particle}. In \cite{chen2013particle}, the fractional Kelvin model successfully characterized this constitutive relationship, expressed as
\begin{equation}
    \frac{{{d^\alpha }\varepsilon }}{{d{t^\alpha }}} + \frac{{{E}}}{\eta }\varepsilon  = \frac{\sigma }{\eta },
\end{equation}
where $\varepsilon$ and $\sigma$ denote strain and stress, respectively, $d^{\alpha}/dt^{\alpha}$ represents the Caputo derivative of fractional order $\alpha$ ($0<\alpha<1$), $E$ is the elasticity modulus, and $\eta$ is the viscosity coefficient. Two types of frozen soil, clay and silt, were tested under confining pressures of $\sigma$=1.11Mpa and with $\sigma$=1.14Mpa, with sample sizes of 56 and 43, respectively. Note that the sampling interval is uneven, and the data contains natural noise. To generate candidate terms for integer- or fractional-order derivatives, data smooth and interpolation at specified positions were necessary. For both datasets, 15 samples were used as the training set, with the remainder used for prediction. Using the trained DNN, which is implemented in PyTorch, the network consists of an input layer (1D$\rightarrow$20D), seven hidden layers (each containing a tanh($\cdot$) activation function), and an output layer (20D$\rightarrow$1D). The model utilizes the Adam optimizer. 400 equidistant reconstructed data points for sparse regression and five auxiliary nodes for each reconstructed points (to compute fractional derivatives) were generated. As illustrated in Figure \ref{fig:creep}, the dataset was effectively smoothened, and auxiliary points required for computing fractional derivatives were arbitrarily prescribed. 
According to the constitutive relationship's nature, the cost function involving the time evolution and candidate library is 

\begin{equation}
    {\cal L(\alpha,\boldsymbol \xi)} =\varepsilon^{(\alpha)}_t - \left[ {\begin{array}{*{20}{c}}
        {1}&{\varepsilon}&{\varepsilon_t}&{\varepsilon_{tt}}&{\varepsilon^2}&{\varepsilon \varepsilon_t}&{\varepsilon_t \varepsilon_t}&{\varepsilon \varepsilon_{tt}}&{\varepsilon_t \varepsilon_{tt}}&{\varepsilon_{tt} \varepsilon_{tt}}\\
        \end{array}} \right] \boldsymbol{\xi},
\end{equation}
and the right-hand side of the equation is $\varepsilon^{(\alpha)}_t$, where the parameter $\alpha$ is unknown and is optimized by the PA, as shown in Algorithm~\ref{tab:pseudocode}.
The discovery results, detailed in Table \ref{tab:fode}, demonstrate that our algorithm successfully identified the fractional Kelvin model structure directly from experimental data, without prior knowledge of the structural form of the differential equations. The parameter estimation error for clay was lower than that for silt (see Table \ref{tab:fode}), likely due to: (1) the sharper transition in the silt stress-strain curve, which provides less information for learning; (2) loss of accuracy in G-J quadrature due to insufficient quadrature points; and (3) the cumulative error of stepwise optimization leads to the deviation of parameters. Improving parameter optimization for FDEs discovery is a potential topic for future work.

\begin{table}[htb]
    \centering
    \caption{Summary of the fractional Kelvin model learned from experimental data}\label{tab:fode}
    \begin{tabular}{@{}llcccccc@{}}
    \toprule
    \multicolumn{1}{c}{Soil type} & Model type & Equation & $\alpha$ & $\eta$ & $E$ & Error\footnotemark[2] \\ 
    \midrule
    \multirow{2}{*}{Clay} 
    & Learned & $\varepsilon^{(0.374)}_t = -2.402 \varepsilon + 8.125$ & 0.374 & 0.137 & 0.328 & 0.151 \\
    & Ground truth \footnotemark[1] & $\varepsilon^{(0.371)}_t = -3.010 \varepsilon + 10.745$ & 0.371 & 0.103 & 0.320 & -- \\
    \addlinespace
    \multirow{2}{*}{Silt} 
    & Learned & $\varepsilon^{(0.448)}_t = -7.344 \varepsilon + 18.955$ & 0.448 & 0.0612 & 0.449 & 0.256 \\
    & Ground truth \footnotemark[1]& $\varepsilon^{(0.562)}_t = -5.681 \varepsilon + 14.910$ & 0.562 & 0.0778 & 0.442 & -- \\
    \bottomrule
    \end{tabular}
    
    \footnotetext[1]{Parameters determined by nonlinear least squares method with predefined model structure.}
    \footnotetext[2]{Average relative error between learned and ground truth parameters.}
    \end{table}

\subsection{Discovering FADE from synthetic data} \label{sec:fade_discovery}

\begin{figure}[htb] 
    \centering
    \includegraphics[width=\textwidth]{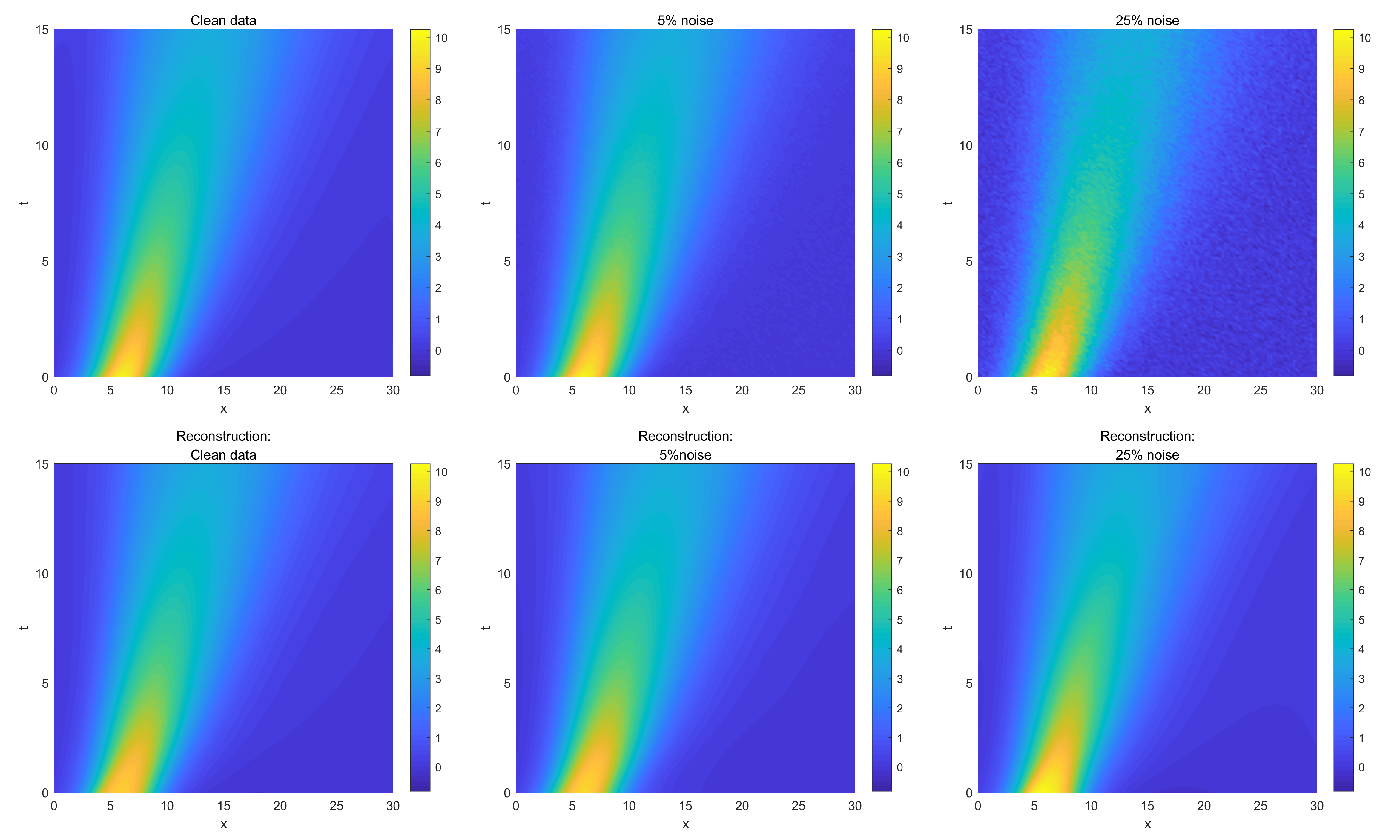}
    \caption{The upper panel illustrates the spatiotemporal distribution of synthetic data, governed by a fractional advection-diffusion equation, under varying noise levels. The lower panel shows the reconstructed data for the upper panel, generated using a DNN.
    }\label{fig:fade_fft}
\end{figure}

In this section, we assess the effectiveness of our method in identifying fractional partial differential equations, with the Fractional Advection-Diffusion Equation (FADE) chosen as the case study. FADE is a parsimonious model that captures anomalous solute diffusion in aquifers \citep{benson2000application}. The mechanisms underlying FADEs vary depending on whether the fractional derivatives relate to time or space: FADEs with time fractional derivatives model delayed transport due to solute retention, while those with space fractional derivatives describe rapid solute displacement along preferential flow paths \citep{zhang2009time}. For generality, we examine FADE with both space and time fractional derivatives. To demonstrate the feasibility of our method in discovering FADE, synthetic data generated through numerical simulation of FADE is employed. The FADE and its initial and boundary conditions in this study are provided as follows:

\begin{equation}
    \centering
    \left\{ \begin{array}{l}
        c_t^{(\alpha )} =  - v{c_x} + Dc_{x}^{(\beta )},\\
        c(x,0) = 10{e^{ - {{[(x - 6)/10]}^2}}},\\
        c(0,t) = c(30,t),
        \end{array} \right.
\end{equation}
where $\alpha=0.8$ and $\beta=1.7$ denote the time and space fractional orders, respectively. A closed-form analytical solution has not yet been found, so we use the fast Fourier transform (FFT) to obtain $E_\alpha[(-iv\kappa+D(i\kappa)^\beta){t^\alpha}]$, where $\kappa$ represents the variable in the frequency domain, and $E_{\alpha}(\cdot)$ denotes the single-parameter Mittag-Leffler function \citep{podlubny1999fractional}, expressed as
\begin{equation}
    {E_{\alpha}}(z) = \sum\limits_{n = 0}^\infty  {\frac{{{z^n}}}{{\Gamma (\alpha n + 1 )}}},
\end{equation}
where $\alpha$ denotes the shape parameter, which represents the fractional order here. An inverse fast Fourier transform (IFFT) is then employed to obtain high-precision semi-analytical solutions, which serve as synthetic data. The cost function involving time evolution and the candidate library is constructed as

\begin{equation} \label{eq:lib_fade}
    {\cal L(\alpha,\beta,\boldsymbol \xi)} = c_t^{(\alpha )} -\left[ {\begin{array}{*{20}{c}}
        {1}&{c}&{c_x}&{c_x^{(\beta)}}&{c_{xxx}}&{c^2}&{c c_x}&{c c_{xx}}&{c c_{xxx}}&{c^2 c_{x}}&{c^2 c_{xx}}&{c^2 c_{xxx}}
        \\
        \end{array}} \right] \boldsymbol \xi,
\end{equation}
and the right-hand-side of equation is the temporal fractional derivative $c^{(\alpha)}_t$. Generally, the advection term of solute transport is characterized by the spatial gradient of concentration, which is a first-order derivative, so we assume the order of the derivative for advection to be an integer. The fractional orders $\alpha$ and $\beta$ are optimized by the DE algorithm, as shown in Algorithm~\ref{tab:pseudocode}.
The distribution of synthetic data at various noise levels and the corresponding reconstructed data are illustrated in Figure \ref{fig:fade_fft}. Note that even clean data also requires reconstruction for computing fractional derivatives. The spatial and temporal scopes are $x=[0,30]$ and $t=[0,15]$, with a resolution of $(x \times t)=(120 \times 150)=18000$, and the time and space intervals are set uniformly to meet the requirements of the FFT-based algorithm. 
In this study, we consider three noise levels, clean (no noise), 5\% uniform noise, and 25\% uniform noise. We randomly select 1000 samples for the training set and 2000 for the validation set, respectively. The proposed neural network architecture is implemented in PyTorch, it consists of an input layer (2D$\rightarrow$20D), five hidden layers (each with Gaussian activation functions having learnable mean and variance), and an output layer (20D$\rightarrow$1D). The model utilizes the Adam optimizer.

The results of FDEs discovery are summarized in Table \ref{tab:fade_noise}. Our method successfully recovers the correct structure of FADE from synthetic data across various noise levels, though the accuracy of parameter estimation (including spatiotemporal fractional orders, velocity and diffusion coefficient) decreases as the noise level increases, as indicated by a rise in mean squared error. This accuracy decline likely results from challenges in data reconstruction under high noise conditions. As noise increases, the fidelity of data reconstruction relative to clean data diminishes, preventing the loss function from accurately reflecting the original parameters of FADE. 


\begin{table}[htb]
    \centering
    \caption{Summary of FADE learned from noisy synthetic data}\label{tab:fade_noise}
    \begin{tabular}{@{}lll@{}}
    \toprule
    Noise level & Learned equation & Error \\ 
    \midrule
    Clean data  & $c^{(0.790)}_t=-1.006c_x+0.501c^{(1.720)}_{x}$ & 0.008 \\
    5\% noise   & $c^{(0.791)}_t=-1.005c_x+0.510c^{(1.667)}_{x}$ & 0.014 \\
    25\% noise  & $c^{(0.804)}_t=-0.946c_x+0.462c^{(1.750)}_{x}$ & 0.041 \\ 
    Ground truth & $c^{(0.8)}_t=-c_x+0.5c^{(1.7)}_{x}$ & -- \\ 
    \bottomrule
    \end{tabular}
    \end{table}

\subsection{Data-driven derivation of Stable Laws}

Previous studies have demonstrated success in data-driven discovery of Lagrangian dynamics \citep{rudy2017data,zhang2020data}. According to the central limit theorem, random walks with a probability density function (PDF) in jump sizes characterized by finite mean and variance (such as Brownian motion) converge to a Gaussian distribution. However, in complex systems where the temporal-spatial distribution of random walker velocities exhibits a long-tail (i.e., power-law) property, this convergence does not hold. Under these conditions, particle trajectories cannot be accurately described by the standard second-order diffusion equation. For example, L\'{e}vy motions, which are memoryless process characterized by finite mean displacement and divergent displacement variance, result in  position densities that converges to the FDE based on the generalized central limit theory \citep{gnedenko1968limit,benson2001fractional}. Consider a random walk for L\'{e}vy motions

\begin{equation} \label{eq:rw}
    Y = X_{1} + X_{2} + \ldots + X_{n},
\end{equation}
where the random variable $Y$ represents the location after $n$ jumps, and $X$ denotes the independent and identically distributed (i.i.d.) jump lengths following an $\alpha$-stable distribution $p(X)=S_\alpha(\beta,\sigma,\mu)$ \citep{levy1954theorie} with parameters $1<\alpha \le 2$, $-1 \le \beta \le 1$, $\sigma>0$ and $\mu$ representing the shape factor, skewness, scale factor and drift, respectively. This random walk describes $n$ jumps of a single particle with a constant time interval $\Delta t$ between two consecutive jumps.  

\begin{figure}[htb] 
    \centering
    \includegraphics[width=\textwidth]{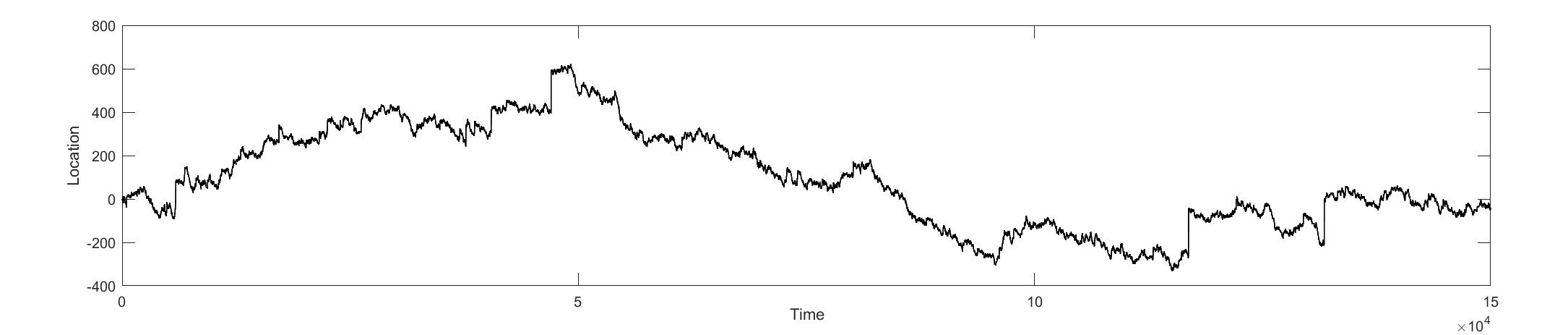}

    \caption{Single particle trajectory of L\'{e}vy motions.}\label{fig:fade_pt}
\end{figure}
  
When $\beta=1$, $\sigma=(\Delta{t}D|\rm{cos}(\pi\alpha/2)|)^{1/\alpha}$ and $\mu=v\Delta{t}$ as $n \rightarrow \infty $, the random variable $Y$ in Equation~\eqref{eq:rw} may converge to a Stable Law (see \ref{sec:eclt}). The PDF of $Y$ is 
\begin{equation} \label{eq:stb_sfade}
    {p_Y}(x,t) = {S_\alpha }(1,{(Dt|{\rm{cos}}(\pi \alpha /2)|)^{1/\alpha }},vt),
\end{equation}
 which corresponds to the solution of the space-fractional diffusion equation (Appendix~\ref{sec:eclt})
\begin{equation} \label{eq:sfade}
    \frac{\partial{c(x,t)}}{{\partial t}} = -v \frac{\partial c(x,t)}{\partial x} + D\frac{{{\partial ^\alpha }c(x,t)}}{{\partial {x^\alpha }}}
\end{equation}
where $v$ represents the mean velocity of forward movement, and $D$ is the effective dispersion coefficient. See Appendix~\ref{sec:eclt} for more details.
Notably, Gulian et al. \cite{gulian2019machine} successfully learned fractional diffusion equations from $\alpha$-stable time series using Gaussian regression, though this approach relies on the availability of the structural form of the equation. The primary objective of this section is to discover the fractional differential equation without relying on any prior information, including its structural form, building on the work of \cite{gulian2019machine}. The candidate library is identical to that in Section~\ref{sec:fade_discovery}.

\begin{figure}[htb] 
    \centering
    \includegraphics[width=\textwidth]{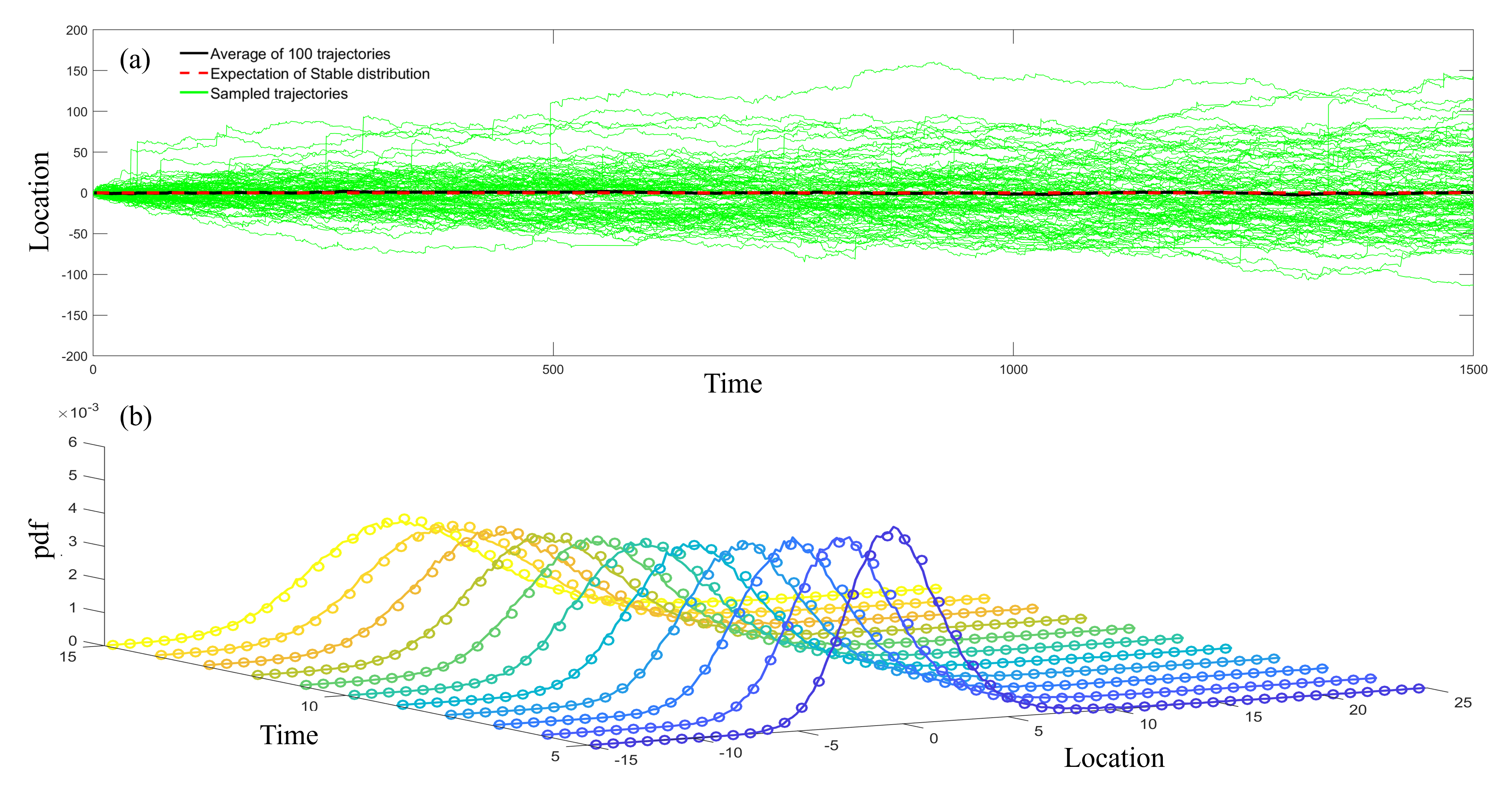}

    \caption{(a) Validation of 100 sampled particle trajectories generated by recovered dynamics. Each of the 100 green lines is a sample path of the stable process generated by recovered dynamics. (b) The distribution of particle numbers generated by original particle trajectory (lines) and learned equation (symbols).}\label{fig:pt_recover}
\end{figure}

We consider a stable time series $S_{1.8}(1,0.66,-0.32)$ with results presented in Table \ref{tab:fde_stable}. The trajectory of this random walker is shown in Figure~\ref{fig:fade_pt}. We begin with a statistical characterization of the particle number density distribution across space-time domains, then employ a DNN to reconstruct the distribution function with noise caused by random number generation. The network architecture configuration is identical to that described in Section \ref{sec:fade_discovery}. The results shown in Table~\ref{tab:fde_stable} and Figure~\ref{fig:pt_recover} demonstrate that a fractional diffusion equation has been successfully identified from the trajectory, the learned fractional orders for time and space are 0.987 and 1.842, respectively, with a diffusion coefficient of 0.475 (Table \ref{tab:fde_stable}). {Note that the temporal order learned by our algorithm closely approximates the true derivative order of one. This indicates that our framework is not limited to extracting FDEs; it can also effectively capture memoryless processes characterized by integer-order temporal derivatives through precise parameter calibration via global optimization.} Figure~\ref{fig:pt_recover} illustrates that our framework can recover the dynamics of the original particle motion.

\begin{table}[htb]
    \centering
    \caption{Summary of the fractional diffusion model learned with $\alpha$-stable time series}\label{tab:fde_stable}
    \begin{tabular}{@{}lll@{}}
    \toprule
     & Learned equation & Error \\ 
    \midrule
    Learned model & $c^{(0.99)}_t=0.48c^{(1.84)}_{xx}$ & 0.029 \\ \\
    Ground truth & $c_t=0.5c^{(1.8)}_{xx}$ & -- \\ 
    \bottomrule
    \end{tabular}
    \end{table}

\section{Discussion}

\subsection{Method comparison}

 To further demonstrate the advancements of our method in discovering FDEs, we compare it with a representative PDE discovery framework, DL-PDE \citep{xu2019dl}, which we consider as one of the most robust algorithms for discovering classical PDEs. 

\subsubsection{Discovery of fractional differential equation}
 The FADE described in Section \ref{sec:fade_discovery} is used as the benchmark for this comparison. 


The candidate library in DL-PDE is similar to that in our method, except it contains only integer-order derivative terms including time evolution $\partial{c}/\partial{t}$ and candidate library
\begin{equation}
    \Theta =\left[ {\begin{array}{*{20}{c}}
        {1}&{c}&{c_x}&{c_{xx}}&{c_{xxx}}&{c^2}&{c c_x}&{c c_{xx}}&{c c_{xxx}}&{c^2 c_{x}}&{c^2 c_{xx}}&{c^2 c_{xxx}}
        \\
        \end{array}} \right].
\end{equation}
Compared to the candidate library in \eqref{eq:lib_fade}, the difference is that DL-PDE replaces the second-order derivative terms with spatial fractional-order derivative terms. The comparison results are shown in Table \ref{tab:comparison_fade}.
From Table \ref{tab:comparison_fade}, when applied to clean data, DL-PDE performs well in capturing specific physics characterized by first- and second-order terms, i.e. advection and local dispersion. The identified orders are the integers closest to the corresponding fractions of the ground truth, and the corresponding coefficients exhibit reasonable accuracy. However, it fails to extract fractional-order derivatives. Under noisy conditions, with 5\% and 25\% uniform noise, DL-PDE is limited to identifying second-order and first-order spatial derivatives, respectively, while demonstrating poor performance in coefficient regression. These findings indicate that DL-PDE is not suitable for discovering FDEs, although it can capture specific characteristics of dispersive transport, such as the mean flow rate (first-order derivative) and deviations from the mean flow rate (second-order derivative).

According to the comparison, our method demonstrates its ability to discover the correct fractional-order of FADE and its coefficients with relatively high accuracy. 

\begin{table}[htb]
    \centering
    \caption{Comparison of FADE results between our method and DL-PDE}\label{tab:comparison_fade}
    \begin{tabular}{@{}lll@{}}
    \toprule
    Noise level & Our framework & DL-PDE \\ 
    \midrule
    Clean data  & $c^{(0.790)}_t=-1.006c_x+0.501c^{(1.720)}_{x}$ & $c_t=-0.964c_x+0.416c_{xx}$ \\
    5\% noise   & $c^{(0.791)}_t=-1.005c_x+0.510c^{(1.667)}_{x}$ & $c_t=1.270c_{xx}$ \\
    25\% noise  & $c^{(0.804)}_t=-0.946c_x+0.462c^{(1.750)}_{x}$ & $c_t=-0.569c_x$ \\ 
    Ground truth & \multicolumn{2}{l}{$c^{(0.8)}_t=-c_x+0.5c^{(1.7)}_{x}$} \\ 
    \bottomrule
    \end{tabular}
    \end{table}

{\subsubsection{Discovery of classical differential equation}}

    To further validate the capability of our framework in discovering standard integer-order differential equations, this section briefly compares its performance with that of DL-PDE. As shown in Table~\ref{tab:comparison_ade}, our method successfully recovers the correct equation structure and parameters, with the identified fractional order closely reaching the true integer order. However, it obtains lower accuracy compared to DL-PDE. This discrepancy is attributed to DL-PDE's exclusive focus on integer-order differential equations, which results in a relatively constrained optimization space. In contrast, our framework accommodates the possibility of identifying non-integer order equations, increasing complexity; furthermore, the optimal linear coefficients may be influenced by the presence of non-integer order derivative terms.
    Notably, our framework exhibits greater noise sensitivity than DL-PDE, as shown by the sharper accuracy decline in Table~\ref{tab:comparison_ade}. This vulnerability is likely attributed to two factors: (1) the higher optimization complexity resulting from a larger parameter space compared to DL-PDE, and 
    (2) Gauss quadrature approximation errors in fractional derivative discretization may cause numerically identified parameters to deviate from theoretical optima. In summary, while our framework exhibits higher sensitivity to noise compared to DL-PDE, it retains acceptable robustness across various noise levels.

\begin{table}[htb]
    \centering
    \caption{{Comparison of FADE results between our method and DL-PDE}}\label{tab:comparison_ade}
    \begin{tabular}{@{}lll@{}}
    \toprule
    Noise level & Our framework & DL-PDE \\ 
    \midrule
    Clean data  & $c^{(0.984)}_t=-1.029c_x+0.210c^{(1.944)}_{x}$ & $c_t=-0.999c_x+0.250c_{xx}$ \\
    5\% noise   & $c^{(0.980)}_t=-1.028c_x+0.212c^{(1.985)}_{x}$ & $c_t=-0.996c_x+0.236c_{xx}$ \\
    25\% noise  & $c^{(0.964)}_t=-1.048c_x+0.148c^{(1.896)}_{x}$ & $c_t=-0.995c_x+0.217c_{xx}$ \\ 
    Ground truth & \multicolumn{2}{l}{$c_t=-c_x+0.250c_{xx}$} \\ 
    \bottomrule
    \end{tabular}
\end{table}
\subsection{The effect of regularization}
Our method employs regularization with a factor $\lambda$ to achieve an effective balance between model parsimony and accuracy in the discovered equations. Particularly, the appropriate setting of $\lambda$ is important for accurately identifying the governing equations (especially in terms of their structure). To determine the suitable magnitudes of $\lambda$ within our framework, we perform a series of numerical experiments on the FADE model under three levels of noise (see also in Section \ref{sec:fade_discovery}). Four different magnitudes of regularization coefficients are considered: $\lambda=0, \lambda=10^{-5}$, $\lambda=10^{-4}$, $\lambda=10^{-3}$, and $\lambda=10^{-2}$. The comparison results of the discovered structures are shown in Table~\ref{tab:comparison_reg}.

\begin{table}[htb]
    \centering
    \caption{The structural form of FADE under different levels of noise and different magnitudes of $\lambda$}\label{tab:comparison_reg}
    \begin{tabular}{@{}llll@{}}
    \toprule
    $\lambda$ & Clean data & 5\% noise & 25\% noise \\ 
    \midrule
    {$0$} & {9 redundant terms} & {9 redundant terms} & {8 redundant terms\footnotemark[1]} \\
    $10^{-5}$ & $c_t^{(0.790)}$, $c_x$, $c^{(1.720)}_x$ & $c_t^{(0.811)}$, $c_x$, $c^{(1.597)}_x$, $c_{xxx}$ & $c_t^{(0.828)}$, $c$, $c_x$, $c^{(1.916)}_x$, $c_{xxx}$, $c^2$, $cc_{xx}$, $cc_{xxx}$ \\
    $10^{-4}$ & $c_t^{(0.796)}$, $c_x$, $c^{(1.641)}_x$ & $c_t^{(0.791)}$, $c_x$, $c^{(1.667)}_x$ & $c_t^{(0.832)}$, 0.0371, $c_x$, $c^2$, $c^2c_{xxx}$ \\
    $10^{-3}$ & $c_t^{(0.795)}$, $c_x$, $c_x^{(1.610)}$ & $c_t^{(0.802)}$, $c_x$, $c^{(1.608)}_x$ & $c_t^{(0.804)}$, $c_x$, $c^{(1.750)}_x$ \\
    $10^{-2}$ & $c_t^{(0.687)}$, $c_x$ & $c_t^{(0.684)}$, $c_x$ & $c_t^{(0.716)}$, $c_x$ \\ 
    Ground truth & \multicolumn{3}{l}{$c_t^{(0.8)}$, $c_x$, $c^{(1.7)}_x$} \\ 
    \bottomrule
    \end{tabular}
    \footnotetext[1]{{Our candidate library consists of 12 terms (3 ground truth and 9 redundant, as shown in Equation~\eqref{eq:lib_fade}); the 9 redundant terms provide complete coverage of meaningless terms. When reduced to 8 redundant terms, only one meaningless term is abandoned.}}
    \end{table}

As shown in Table \ref{tab:comparison_reg}, although smaller $\lambda$ may increase the accuracy of coefficient identification (i.e., fractional order), some redundant terms are identified due to the overfitting of sparse regression when $\lambda$ is small. {First, the absence of regularization ($\lambda=0$) leads to overfitting in equation discovery, resulting in many redundant terms in the solution. Consequently, the unregularized results fail to capture the underlying physical mechanism and are excluded from subsequent analyses.} When $\lambda=10^{-5}$, the discovered equation under 5\% noise includes the term $c_{xxx}$, which is absent in the ground truth. On the other hand, increasing $\lambda$ appropriately helps preserve model parsimony, but a large value of $\lambda$ makes the model overly parsimonious, resulting in the lack of key terms in the discovered equation. For instance, when $\lambda=10^{-2}$, the discovered equation lacks the spatial fractional derivative term $c^{(1.7)}_x$ compared to the ground truth. 

Furthermore, the suitable range of $\lambda$ varies with the noise intensity.
For instance, with clean data, $\lambda=10^{-5}$ yields a discovered equation consistent with the ground truth, while under 5\% and 25\% noise, it contains one and five redundant terms, respectively. Similarly, when $\lambda=10^{-4}$, the discovered equation is consistent with the ground truth for clean data and 5\% noise, but under 25\% noise, it contains three false terms.  This indicates that lower noise levels allow for a broader range of optimal $\lambda$.
In summary, $\lambda=10^{-3}$ is the most practical choice for {FDEs} discovery, as it successfully identifies the correct structural form of the FADE and its parameters under various noise levels.






\section{Conclusions}
In this study, we present a stepwise framework for identifying fractional differential equations directly from data. 
We validate our approach by recovering commonly-used FDEs across three scenarios: experimental data of frozen soil creep behavior, synthetic data of solute transport in aquifers, and a single particle trajectory of L\'{e}vy motion. These case studies demonstrate that our algorithm effectively extracts practical FDEs, showing robustness across various levels of natural and artificial noise. 
The algorithm integrates both fractional-order and integer-order derivatives into the candidate library, broadening its scope of applicability. Moreover, memoryless dynamics can be captured by calibrating the fractional order to approximate integer values.
To further validate the advancements of our framework in identifying FDEs, a comparative analysis with a representative existing method, DL-PDE, is conducted. Additionally, the applicable range of the regularization factor for different levels of data noise is analyzed. The results indicate that as noise intensity decreases, the range of suitable regularization values broadens, and $\lambda=10^{-3}$ falls within the optimal range across different noise levels.

Several avenues for further investigation remain. First, the proposed framework is relatively time-consuming, requiring multiple steps to optimize different parameter groups. Future work could focus on developing a more efficient algorithm to reduce the computational cost of the proposed framework. Additionally, the efficiency and robustness of equation identification can be largely improved by effectively utilizing available prior knowledge or enhancing the algorithm.
  Second, besides the FDEs in this study, there is a wide spectrum of parsimonious models (maybe uncovered) aimed at capturing different mechanisms that play important roles in the dynamics of complex systems. Discovering these models could be a promising direction for future work. To achieve this goal, a large, closed library that encompasses all possible candidate terms for describing the dynamics of complex systems is required, which is however unrealistic. Improving the flexibility of our algorithm by reducing the reliance on an overcomplete candidate library offers promising solutions to this challenge, such as adopting expandable libraries \citep{xu2020dlga} or using open-formed algorithms \citep{chen2022symbolic}. {Third, the current framework's reliance on spatiotemporal data for derivative computation and its inefficiency with sparse observations (e.g., 1D time series common in subsurface systems) present key limitations. Promising avenues to address these challenges include developing coarse-grained reduced-order models to bridge the observation-dimensionality gap \cite{brunton2024promising}, and using Neural fractional differential equations for their unique capability to capture memory-dependent dynamics from limited time-series data while maintaining extrapolation accuracy \cite{coelho2025neural}.}

\section*{Acknowledgement}
This work (except YZ) is financially supported by the National Key Research and Development Program of China (No. 2024YFF1500600), the Natural Science Foundation of Ningbo of China (No. 2023J027), the National Natural Science Foundation of China under Grant (No. 12171404), as well as by the High Performance Computing Centers at Eastern Institute of Technology, Ningbo, and Ningbo Institute of Digital Twin. Y.Z. was partially funded by the National Science Foundation (Grant 2412673), United States. The results of this study do not reflect the view of the funding agencies.

\section*{Data availability}
{The codes and datasets for this research is available at: \url{https://github.com/yxn1019/FDE_discovery.git}. }

\appendix

\section{{Brief Review of Neural Fractional Differential Equation (Neural FDE) Literature and Positioning of Our Work}}
{ In recent years, significant research has focused on Neural Fractional Differential Equations (Neural FDEs), extending the Neural ordinary differential equation (Neural ODE) framework by incorporating fractional-order derivatives to better model systems with memory effects and long-range dependencies. Most of these works highlight the potential of deep learning to handle complex dynamics but primarily focus on black-box approximation rather than interpretable model discovery. Below, we summarize major contributions in this area and clarify how our framework complements and extends this body of work.

Coelho et al. \cite{coelho2024tracing} introduced the foundational Neural FDE architecture, integrating fractional derivatives into Neural ODEs. Their method jointly trains a neural network to model system dynamics and learns the fractional derivative order, enabling improved extrapolation performance, especially for systems with inherent memory.
Coelho et al. \cite{coelho2024neural} explored the non-uniqueness problem in Neural FDEs, analyzing how initialization and optimization of the fractional order can lead to multiple valid solutions, and discussing implications for interpretability and physical consistency.
Kang et al. \cite{kang2024unleashing} proposed FROND, coupling Graph Neural Networks (GNNs) with fractional-order dynamics. Their study demonstrated that fractional derivatives improve the robustness of GNNs to adversarial attacks, extending the utility of Neural FDEs to graph-based data.
Zhang et al. \cite{zhang2024fde} introduced FDE-Net, leveraging fractional dynamics to design densely connected neural architectures for single-image super-resolution tasks, showing how fractional calculus can inspire network design, not just system modeling.
Zimmering et al. \cite{zimmering2024optimising} focused on optimizing solver efficiency by developing a faster Predictor-Corrector implementation for Neural FDEs, achieving significant improvements in both speed and accuracy, and benchmarking against Neural ODEs for fair comparison.
Coelho et al. \cite{coelho2025neural} offered a deeper theoretical foundation and expanded experiments. They emphasized the effectiveness of learning alongside the neural parameters, showing enhanced accuracy for synthetic and real-world datasets.
Cui et al. \cite{cui2025neural} extended the framework to Neural Variable-Order FDEs (NvoFDEs), allowing the fractional order to vary dynamically with system states, offering improved flexibility and adaptability for systems where memory effects are spatially or temporally heterogeneous.
Kang et al. \cite{kang2025efficient} proposed an adjoint back propagation method to tackle the high computational cost associated with Neural FDEs, introducing an efficient way to train these models without sacrificing performance - addressing a main limitation of earlier works.

\subsection*{Positioning and Distinction of Our Work}
While the Neural FDE literature has made substantial strides in approximating the behavior of systems governed by FDEs, these methods primarily function as black-box predictors. They are highly effective for forward simulations and extrapolation but do not yield explicit governing equations, limiting interpretability and scientific insight.
Our framework departs from this paradigm in several fundamental ways:

(1) White-Box Model Discovery:

The principal goal of our method is to discover explicit, closed-form FDEs directly from data. This white-box approach provides interpretable models that offer deep insight into the underlying physical processes, enabling validation, theoretical exploration, and further analytical study—capabilities that are absent in most Neural FDE frameworks.

(2) Hybrid Architecture:

Although we incorporate DNNs as surrogate models for tasks like data denoising and reconstruction (which leverage the strengths of black-box methods), the core discovery process relies on sparse regression and global optimization. This ensures that the ultimate output is a human-readable, physically meaningful equation, bridging the gap between black-box learning and traditional scientific modeling.

(3) Complementary Contribution to Neural FDEs:

By situating our work within the broader context of Neural FDE research, we address a crucial need: moving beyond mere prediction to explicit model identification. Our framework is particularly valuable in scientific and engineering disciplines where uncovering the true governing equations is essential for validation, regulation, or mechanistic understanding—complementing Neural FDEs’ strength in predictive modeling.

In summary, while Neural FDEs have expanded the reach of machine learning into fractional dynamics with notable success, our framework extends this progress by offering a transparent, interpretable, and physically grounded alternative, fulfilling a distinct and complementary role in the landscape of data-driven modeling.
}

\section{Gaussian-Jacobi quadrature}\label{sec:gj}
Due to their nonlocal and singular nature, approximating fractional derivatives presents numerical challenges. Nonlocality increases computational cost, particularly at large time or space scales, as it requires dense mesh nodes. To address this issue, we employ Gaussian-Jacobi (G-J) quadrature, an effective method for integrands with endpoint singularities.
G-J quadrature is a refined numerical integration technique that provides accurate approximations for integrals with power-law weight functions like $(1-x)^\lambda (1+x)^\mu$. The general quadrature formula is as follows

\begin{equation}\label{eq:gj_rule}
    \int_{ - 1}^{ + 1} {f(x ){\rho ^{(\mu {\rm{,}}\lambda )}}} (x )dx  = \sum\limits_{i = 1}^N {{\omega _i}f({\xi _i})},
\end{equation}
{where $\rho^{(\mu,\lambda)}=(1-x)^{\lambda} (1+x)^{\mu}$ denotes the singular kernel function, and $\omega_i$ represents weight factors determined by the G-J rule, as referenced in \cite{shen2011spectral,li2019theory}. $N$ denotes the prescribed number of quadrature nodes. The quadrature nodes, denoted by $\xi$ and corresponding weight factors $\omega_i$, can be computed using either the Golub-Welsch algorithm \cite{golub1969calculation} or directly via the SciPy Python package.}
 The G-J rule efficiently computes the fractional integral of any smooth function. By substituting $\epsilon(\tau) = (t-a)(1 + \tau)/2$, Equation \eqref{eq:frac_int} can be rewritten as

\begin{equation}\label{eq:gj_frac}
    I_a^\gamma f(t) = \frac{{{{(t - a)}^\gamma }}}{{\Gamma (\gamma )}}\int_{-1}^1 {{{(1 + \tau )}^{\gamma  - 1}}f[\epsilon(t,\tau)]} d\tau,
\end{equation}
which can be regarded as the G-J quadrature \eqref{eq:gj_rule} with $\mu = 0$ and $\lambda = \gamma-1$. The fractional derivatives can then be obtained using Equations \eqref{eq:rl_def} and \eqref{eq:cap_def}. The G-J rule's ability to handle singular integrals is a significant advantage in solving FDEs, enabling effective and accurate solutions. This approach allows us to embed the fractional derivative term into the candidate library, resulting in solving this system, we obtain approximate solutions that accurately capture the behavior of FDEs, even in the presence of singularities.

\begin{figure}[htb]
    \centering
    \includegraphics[width=0.45\textwidth]{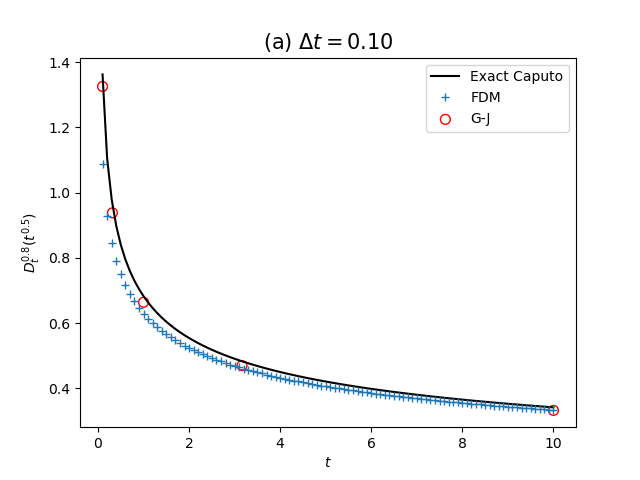}
    \includegraphics[width=0.45\textwidth]{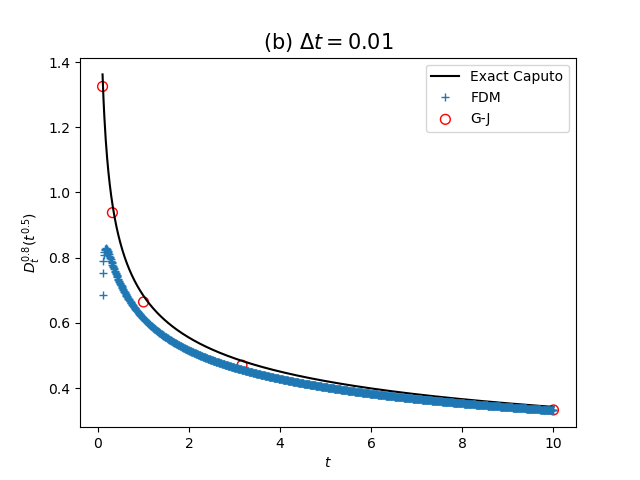}
    \caption{Comparison of differentiation strategies, finite difference method (FDM) versus Gaussian-Jacobi quadrature (G-J) for fractional derivatives.}
    \label{fig:gj}
\end{figure}

We examine the example $\frac{d^{4/5}t^{1/2}}{dt^{4/5}}$, using the Caputo definition of the fractional derivative. The exact solution is given by $\Gamma(1.5)t^{-0.3}/\Gamma(2.3)$, highlighting the singularity at the origin. We compare the numerical performance of the finite difference method (FDM) \citep{lin2007finite} and G-J quadrature. As shown in Figure \ref{fig:gj}, the inherent singularity in fractional derivatives results in insufficient accuracy for FDM, and precision does not improve (see Table~\ref{tab:time}) as the temporal grid $\Delta t$ decreases from 0.1 (Fig. \ref{fig:gj}a) to 0.01 (Fig. \ref{fig:gj}b). Accuracy is particularly low near the origin. In contrast, the G-J quadrature method achieves a high precision with only five quadrature points, and performs well at the origin. Moreover, with higher accuracy, G-J quadrature is more efficient than the FDM (Table \ref{tab:time}). Thus, G-J quadrature outperforms FDM, especially in scenarios where singularities exist.
 
\begin{table}[htb]
    \centering
    \caption{Time cost and precision of Gaussian-Jacobi quadrature and finite difference method}\label{tab:time}
    \begin{tabular}{@{}lcc@{}}
    \toprule
    Method & CPU time (s) & $l_2$-error \\ 
    \midrule
    FDM ($\Delta t=0.1$) & 0.0081 & 17.35 \\
    FDM ($\Delta t=0.01$) & 0.78 & 163.19 \\
    G-J quadrature\footnotemark[1] & 0.00028 & 1.85 \\
    \bottomrule
    \end{tabular}
    
    \footnotetext[1]{Number of G-J nodes is 5 (independent of time interval $\Delta t$)}
    \end{table}

\section{Stable Law} \label{sec:eclt}
\subsection{Stable distribution}

The probability density function of a stable distribution has no explicit formula; it is represented by its Fourier transform \citep{feller1991introduction}

\begin{equation} \label{eq:stb}
    \hat p(k) = \exp \left[ { - |k{|^\alpha}{\sigma ^\alpha }(1 + {\mathop{\rm i}\nolimits} \beta {\mathop{\rm sign}\nolimits} (k)\tan (\pi \alpha /2) - \mu {\mathop{\rm i}\nolimits} k)} \right]
\end{equation}
where $0<\alpha \le 2$, $\sigma>0$, $-1 \le \beta \le 1$, and $\mu$ represent the shape, scale, skewness and drift respectively. Setting $\sigma=(C|cos(\pi \alpha/2)|)^{1/\alpha}$ and $\beta=p-q$, where $C$ is a positive parameter and $p+q=1$, we obtain the equivalent form \citep{benson2001fractional} 

\begin{equation} \label{eq:stb_fade}
    \hat p(k) = \exp \left[ {qC{{( - {\mathop{\rm i}\nolimits} k)}^\alpha } + pC{{({\mathop{\rm i}\nolimits} k)}^\alpha } - \mu {\mathop{\rm i}\nolimits} k} \right].
\end{equation}
Particularly, when $\beta=1$ (i.e., $p=1$ and $q=0$), $\mu=v$ and $C=Dt$, the function in $\hat{p}(k)$ in \eqref{eq:stb_fade} is equal to the function \eqref{eq:stb_sfade}, which is the solution to the space fractional advection-diffusion \eqref{eq:sfade} under the application of the Fourier transform.
\subsection{L\'{e}vy motion}
Now consider the Langevin equation for single particle motion
\begin{equation} \label{eq:langevin}
    dX(t)=Vdt+BdL_{\alpha}(t).
\end{equation}
Following the detailed study by \cite{yong2006using}, when $V=v$, $B=[D|\rm{cos}(\pi \alpha/2)|]^{1/\alpha}$ and $dL_\alpha(t)=(dt)^{1/\alpha}S_\alpha(\beta=1,\sigma=1,\mu=0)$, the particle number density for Equation~\eqref{eq:langevin} satisfies the space-fractional advection-diffusion equation \eqref{eq:sfade}. By coarse-graining the Langevin equation \eqref{eq:langevin} with $\Delta{t} \approx dt$, we obtain
\begin{equation} \label{eq:recursion}
    X(t_{i+1})=X(t_i) + V \Delta{t} + B\Delta{t}^{1/\alpha} S_\alpha(\beta=1,\sigma=1,\mu=0),
\end{equation}
where $\Delta{t}=t_{i+1}-t_i$ and $i=1,2,...,n$. According to the parameterization law by \cite{nolan1998parameterizations}, $X(t)$ follows 

\begin{equation} \label{eq:cg}
    X({t_{i + 1}}) = X(t_i) + {S_\alpha }(1,\sigma,v\Delta t),
\end{equation}
where $\sigma={(\Delta tD|{\rm{cos}}(\pi \alpha {\rm{/2}}){\rm{|}})^{{\rm{1/}}\alpha }}$. The recursion for the random variable $X$ in Equation~\eqref{eq:cg} represents the L\'{e}vy motion \eqref{eq:rw}, whose density converges to the solution of the space FADE \eqref{eq:sfade}. Thus, the L\'{e}vy motion \eqref{eq:rw} can be considered a coarse-grained representation of the Markov process \eqref{eq:langevin}, whose scaling limit is the space FADE \eqref{eq:sfade}.

\bibliographystyle{unsrt}
\bibliography{fPINN}
\end{document}